\documentclass[aps,groupedaddress,showpacs]{revtex4}
\usepackage[dvips]{graphicx}

\def \V {\rule{0pt}{2.5ex}}
\def \D {\rule[-1ex]{0pt}{0pt}}

\begin{document}
\title{Possible alterations of the gravitational field in a superconductor}
\author{G.A. Ummarino}
\affiliation{$INFM-$Dipartimento di Fisica, Politecnico di
Torino, Corso Duca degli Abruzzi, 24-10129 Torino, Italy\\ E-mail
address: ummarino@polito.it}

\date{\today}

\begin{abstract}
In this paper I calculate the possible alteration of
the gravitational field in a superconductor by using the
time-dependent Ginzburg-Landau equations (TDGL). I compare the
behaviour of a high-T$_{c}$ superconductor (HTCS) like
YBa$_{2}$Cu$_{3}$O$_{7}$ (YBCO) with a classical low-T$_{c}$
superconductor (LTCS) like Pb. Finally, I discuss what values of
the parameters characterizing a superconductor can enhance the
reduction of gravitational field. 
\end{abstract}

\maketitle


There is no doubt that the interplay between gravitational field
and superconductivity is a very intriguing field of research,
whose theoretical study has been involving many researchers for a
long time\cite{refe0a,refe0b,refe0c,refe0d,refe1a}. Eight years
ago, E.~Podkletnov and R.~Nieminem declared the achievement of an
experimental evidence for a gravitational shielding due to a
rotating high-$T_{\text{c }}$ superconductor. After their
announcement, other groups tried to repeat the experiment but they
obtain controversial results \cite{refe2a,refe3a,refe4a}, so that
at the present moment the question is still open.

In 1996, G.~Modanese interpreted the results by Podkletnov and
Nieminem in the frame of quantum theory of General Relativity
\cite{refe5a} but the complexity of the formalism he used makes it
very difficult to extract quantitative predictions.

In a very recent paper, M.~Agop~\cite{refe6a} and collaborators
wrote generalized Maxwell equations that simultaneously treat weak
gravitational and electromagnetic fields. In the weak field
approximation, the Einstein equations \cite{refe7aa},
\[
R^{\mu\nu}-\frac{1}{2}R g^{\mu\nu}=\frac{-8\pi G
}{c^{4}}T^{\mu\nu}  \nonumber
\] lead to the following equations
\cite{refe7a}, formally similar to Maxwell's:
\begin{equation}
{\bf \nabla }\cdot {\bf E}_{g}=-4\pi G\rho _{g}  \label{1}
\end{equation}
\begin{equation}
{\bf \nabla }\cdot {\bf B}_{g}=0  \label{2}
\end{equation}
\begin{equation}
{\bf \nabla }\times {\bf E}_{g}=-\frac{\partial {\bf
B}_{g}}{\partial t} \label{3}
\end{equation}
\begin{equation}
{\bf \nabla }\times {\bf B}_{g}=-\frac{4\pi G}{c^{2}}{\bf j}_{g}+\frac{1}{%
c^{2}}\frac{\partial {\bf E}_{g}}{\partial t}  \label{4}
\end{equation}
where ${\bf E}_{g}$ and ${\bf B}_{g}$ are the gravitoelectric and
gravitomagnetic field respectively, ${\bf j}_{g}$ is the mass
current density vector such that ${\bf j}_{g}={\bf v}%
\rho _{g}$, ${\bf v}$ is the velocity, $\rho _{g}$ is the mass
density. Obviously G is the Newton's constant and $c$ is the speed
of light in vacuum. As in the electromagnetic case, it is possible
to define a gravitational permitivity $ \varepsilon_{g}=1/4\pi G$
and a gravitational permeability $\mu_{g}=4 \pi G/c^{2}$ of the
vacuum. For example on the surface of the earth ${\bf E}_{g}$ is
simply the Newtonian gravitational acceleration and ${\bf B}_{g}$
is related to angular momentum interactions \cite{refe6a,refe7a}.
 Then, they  defined
generalized electric field, magnetic field, scalar and vector
potentials containing both an electromagnetic and a gravitational
term, in the following way:
${\bf E=E}_{e}{\bf +}\frac{m}{e}{\bf E}_{g}$ ; ${\bf B=B}_{e}{\bf +}\frac{m}{%
e}{\bf B}_{g}$; ${\bf \phi =\phi }_{e}{\bf +}\frac{m}{e}{\bf \phi
}_{g}$ and ${\bf A=A}_{e}{\bf +}\frac{m}{e}{\bf A}_{g}$ where $m$
and $e$ are the electronic mass and charge  and the subscripts $e$
and $g$ mean `electromagnetic' and `gravitational' respectively.
The generalized Maxwell equations then become~\cite{refe6a}:
\begin{equation}
{\bf \nabla }\cdot {\bf E}=\left( -4\pi G+\frac{1}{\varepsilon
_{0}}\right) \rho  \label{5}
\end{equation}
\begin{equation}
{\bf \nabla }\cdot {\bf B}=0  \label{6}
\end{equation}
\begin{equation}
{\bf \nabla }\times {\bf E}=-\frac{\partial {\bf B}}{\partial t}
\label{7}
\end{equation}
\begin{equation}
{\bf \nabla }\times {\bf B}=\left( \frac{-4\pi G}{c^{2}}+\mu
_{0}\right) {\bf j}+\frac{1}{c^{2}}\frac{\partial {\bf
E}}{\partial t} \label{8}
\end{equation}
\noindent{where the relations: $\rho _{g}=\frac{m}{e}\rho $ and{\bf \ }${\bf j}_{g}=%
\frac{m}{e}{\bf j}$ have been used and $\epsilon_{0}$ and
$\mu_{0}$ are the electric permitivity and magnetic permeability
in the vacuum. } They also wrote the two generalized London
equations \cite{refe6a}
\begin{eqnarray}
{\bf E} &=&\left( 1/\rho \right) \left( \partial {\bf j/\partial
}t\right) \label{9} \\ {\bf B} &=&\left( -1/\rho \right) {\bf
\nabla }\times {\bf j}  \nonumber
\end{eqnarray}
and so they could define the generalized penetration depth
\begin{equation}
\lambda =\frac{\lambda _{g}\lambda _{e}}{\sqrt{\lambda
_{g}^{2}-\lambda _{e}^{2}}}\simeq \lambda _{e}  \label{10}
\end{equation}
where $\lambda _{e}=\left[ m/(\mu _{0}e^{2}n)\right] ^{1/2}$, $\lambda _{g}=%
\left[ c^{2}/(4\pi Gmn)\right] ^{1/2},$ $n$ is the density of superelectrons
and $\lambda _{g}/\lambda _{e}\simeq 10^{21}.$

For simplicity, I will study the case of an \emph{isotropic}
superconductor, in the gravitational field of the earth, in the
absence of an electromagnetic field, thus taking $\bf{E_{e}=0}$
and $\bf{B_{e}=0}$. ${\bf B}_{g}$ in the solar system is very
small \cite{refe11a} therefore, ${\bf E=} \frac{m}{e}{\bf E}_{g}$
and ${\bf B=0}$. Moreover the gravitational effects of ${\bf
j}_{g}$ are insignificant (${\bf j}_{g}$ is related to
gravitational effect of the superconductor) and so I assume ${\bf
j}_{g}=0$. I have also ${\bf \phi =}\frac{m}{e}{\bf \phi }_{g}$
and ${\bf A=}\frac{m}{e}{\bf A}_{g}$. This situation isn't
analogous of the Meissner effect but, rather, to the case of a
superconductor in a electric field. Since the gravitoelectric
fied is formally analogous to electric field I will use the
time-dependent Ginzburg-Landau equations (TDGL)
\cite{refe8a,refe9a,refe10a}, which, in the Coulomb gauge ${\bf
\nabla }\cdot {\bf A}=0$, are written in the form:
\begin{equation}
\frac{\hbar^{2}}{2mD}\left( \frac{\partial }{\partial
t}+\frac{2ie}{\hbar}\phi
\right) \psi - a \psi +b\left| \psi \right| ^{2}\psi+\frac{1}{2m}(i\hbar%
{\bf \nabla }+\frac{2e}{c}{\bf A})^{2}\psi =0 \label{11}
\end{equation}
\begin{equation}
{\bf \nabla \times \nabla \times A-\nabla \times H=}\frac{-4\pi \sigma }{c}(%
\frac{1}{c}\frac{\partial {\bf A}}{\partial t}+{\bf \nabla }\phi )+\frac{%
4\pi }{c}\left[ \frac{e\hbar}{mi}(\psi ^{\ast }{\bf \nabla }\psi -\psi {\bf %
\nabla }\psi ^{\ast })-\frac{4e^{2}}{mc}\left| \psi \right|
^{2}{\bf A}\right]  \label{12}
\end{equation}
where $D$ is the diffusion coefficient, $\sigma $  is the
conductivity in the normal phase, \textbf{H} is the applied field,
$a(T)=a_{0}(T-T_{c})$ and $b(T)\equiv b(T_{c})$ where $a_{0}$ and
$b$ are the positive constants and $T_{c}$ is the critical
temperature of the superconductor.
The boundary and initial conditions are:
\begin{equation}
\left.
\begin{array}{c}
(i\hbar{\bf \nabla }\psi +(2e/c){\bf A}\psi )\cdot {\bf n}=0\text{
} \\ {\bf \nabla \times A\cdot n=H\cdot n} \\ {\bf A\cdot n=}0
\end{array}
\right\} \text{ on }\partial \Omega \times (0,t) \hspace{20mm}
\left.
\begin{array}{c}
\psi (x,0)=\psi _{0}(x) \\
{\bf A}(x,0)={\bf A}_{0}(x)
\end{array}
\right\} \text{ on } \Omega\label{boundary}
\end{equation}
where  $\partial \Omega $ is the boundary of a smooth and simply
connected domain $\Omega $ in R$^{\text{n}}$.
 In order to write
equations \ref{11},\ref{12} in a dimensionless form, the following
quantities can be defined:
\begin{equation}
\Psi ^{2}(T)=\frac{\left| a(T)\right| }{b};\text{
}H_{c}(T)=\sqrt{\frac{4\pi \mu _{0}\left| a(T)\right|
^{2}}{b}}=\frac{h/2e}{2\surd2\pi\lambda(T)\xi(T)} \label{16}
\end{equation}
\begin{equation}
\xi (T)=\frac{h}{\sqrt{2m\left| a(T)\right| }};\text{ }\lambda (T)=\sqrt{%
\frac{bmc^{2}}{4\pi \left| a(T)\right| e^{2}}} \label{17}
\end{equation}
\begin{equation}
\kappa =\lambda (T)/\xi (T),\text{ }\tau(T) =\lambda
^{2}(T)/D,\text{ }\eta =4\pi \sigma D/(\varepsilon
_{0}c^{2})\label{18}
\end{equation}
\noindent{where $\lambda (T)$, $\xi (T)$ and $H_{c}(T)$ are the
penetration depth, the coherence length and the thermodynamic
field.}

The dimensionless quantities are then:
\begin{equation}
x^{\prime }=x/\lambda,\text{ }t^{\prime }=t/\tau ,\text{ }\psi
^{\prime }=\psi /\Psi  \label{19}
\end{equation}
\begin{equation}
{\bf A}^{\prime }={\bf A}\kappa /(\sqrt{2}H_{c}\lambda),\text{
}\phi
^{\prime }=\phi \kappa /(\sqrt{2}H_{c}D),\text{ }H^{\prime }=H\kappa /(%
\sqrt{2}H_{c}).  \label{20}
\end{equation}

Inserting the eq.~\ref{19},\ref{20} in eqs.~\ref{11},\ref{12} and
dropping the prime gives the dimensionless TDGL equations
\cite{refe8a}
in a bounded, smooth and simply connected domain $\Omega $ in R%
$^{n}$:
\begin{equation}
\frac{\partial \psi }{\partial t}+i\phi \psi +\kappa ^{2}(\left|
\psi \right| ^{2}-1)\psi +(i{\bf \nabla }+{\bf A})^{2}\psi =0
\label{21}
\end{equation}
\begin{equation}
\eta (\frac{\partial {\bf A}}{\partial t}+{\bf \nabla }\phi )+\frac{1}{2}%
i(\psi ^{\ast }{\bf \nabla }\psi -\psi {\bf \nabla }\psi ^{\ast
})+\left| \psi \right| ^{2}{\bf A+\nabla \times \nabla \times
A-\nabla \times H}=0 \label{22}
\end{equation}

The boundary and initial conditions (\ref{boundary}) become, in
the dimensionless form:
\begin{equation}
\left.
\begin{array}{c}
(i{\bf \nabla }\psi +{\bf A}\psi )\cdot {\bf n}=0\\ {\bf \nabla
\times A\cdot n=H\cdot n} \\ {\bf A\cdot n=}0
\end{array}
\right\} \text{ on }\partial \Omega \times (0,t) \left.
\hspace{20mm}
\begin{array}{c}
\psi (x,0)=\psi _{0}(x) \\ {\bf A}(x,0)={\bf A}_{0}(x)
\end{array}
\right\} \text{ on }\Omega \label{boundary3}
\end{equation}

Our superconductor is immersed in the gravitational field of the
earth which is very weak and approximately constant. So $\phi
=-g^{\ast }x$ where $g^{\ast }=\lambda(T)\kappa
mg/(\sqrt{2}eH_{c}(T)D)\ll 1$ and $g$ is the gravity acceleration.
 The corrections to $\phi$ in the superconductor are of the second
 order in $g^{\ast }$ and therefore they aren't considered here.

Now I search for a solution of the form:
\begin{eqnarray}
\psi (x,t) &=&\psi _{0}(x,t)+g^{\ast }\gamma (x,t)  \label{26} \\
A(x,t) &=&0+g^{\ast }{\bf \beta }(x,t)  \nonumber \\ \phi(x)
&=&-g^{\ast }x  \nonumber
\end{eqnarray}

At order zero in $g^{\ast}$, eq.(\ref{21}) gives:
\begin{equation}
\frac{\partial \psi _{0}(x,t)}{\partial t}+\kappa ^{2}(\left| \psi
_{0}(x,t)\right| ^{2}-1)\psi (x,t)-\frac{\partial ^{2}\psi _{0}(x,t)}{%
\partial x^{2}}=0  \label{27}
\end{equation}
with the conditions: \vspace{-3mm}
\begin{eqnarray}
\psi _{0}(x,t =0)&=&0 \nonumber
\\ \psi _{0}(x =0,t)&=&0 \label{28}
\\ \psi _{0}(x =L,t)&=&0 \nonumber
\end{eqnarray}
\noindent{where $L$ is the length of the superconductor and $t=0$
is the instant when the material undergoes the transition to the
superconducting state.}

The static classical solution of eq.~\ref{27} is:
\begin{equation}
\psi _{0}(x,t)\equiv \psi_{0}(x)=\left\{ \tanh \left[ \kappa
x/\sqrt{2}\right] -\tanh \left[ \kappa (x-L)/\sqrt{2}\right]
-\tanh \left[ \kappa L/\sqrt{2}\right] \right\} /\tanh \left[
\kappa L/\sqrt{2}\right]. \label{29}
\end{equation}

At the first order in $g^{\ast}$ one obtains from equation
(\ref{21}):
\begin{equation}
\frac{\partial \gamma (x,t)}{\partial t}-\frac{\partial ^{2}\gamma (x,t)}{%
\partial x^{2}}+\kappa ^{2}(3\left| \psi _{0}(x)\right| ^{2}-1)\gamma
(x,t)=ix\psi _{0}(x)  \label{30}
\end{equation}
with the conditions:
\begin{eqnarray}
\gamma (x,t =0)&=&0  \label{31} \\ \gamma (x =0,t)&=&0  \nonumber
\\ \gamma (x =L,t)&=&0  \nonumber
\end{eqnarray}

The equation at order one for the vector potential is
\begin{equation}
\eta \frac{\partial \beta (x,t)}{\partial t}+\left| \psi
_{0}(x)\right| ^{2}\beta (x,t)+ J(x,t)-\eta =0  \label{32}
\end{equation}
with the constraint
\begin{equation}
\beta (x,t=0)=0  \label{33}
\end{equation}

Note that the second-order spatial derivative of $\beta$ does not
appear in eq.(\ref{32}). This is due to the fact that, in one
dimension, ${\bf \nabla }^{2}A=\frac{\partial }{\partial x}{\bf
\nabla \cdot A}$ and therefore, in the Coulomb gauge, ${\bf \nabla
\times \nabla\times A=\nabla (\nabla \cdot A)-\nabla }^{2}A=0$.
The quantity $J(x,t)$ which appears in eq.~\ref{32} is given by:
\begin{equation}
J(x,t)=\frac{1}{2}\left[ \psi _{0}(x)\frac{\partial }{\partial x}
{\mathrm Im} \gamma (x,t)-{\mathrm Im}\gamma (x,t)\frac{\partial
}{\partial x}\psi _{0}(x)\right]  \label{34}
\end{equation}

The solution of eq.~\ref{32} is
\begin{equation}
{\bf \beta }(x,t)=\frac{\eta }{\left| \psi _{0}(x)\right|
^{2}}\left[ 1-\exp (-\left| \psi _{0}(x)\right| ^{2}t/\eta
)\right] -\frac{\exp (-\left| \psi _{0}(x)\right| ^{2}t/\eta
)}{\eta }\int_{0}^{t}dt J(x,t)\exp (\left| \psi _{0}(x)\right|
^{2}t/\eta )  \label{35}
\end{equation}

Now, we have an expression for $\psi_{0}(x,t)$ (eq.~\ref{29}), and
an expression for $\beta(x,t)$ as a function of $\gamma(x,t)$
(eq.~\ref{35}). Therefore, we can insert these expressions in
eq.~\ref{26} and obtain both $\psi(x,t)$ and $\mathrm{\bf
A}(x,t)$ as functions of $\gamma(x,t)$. Using the relation:
\begin{equation}
{\bf E}_{g}=-{\bf \nabla \phi -}\frac{\partial {\bf A}}{\partial
t} \label{36}
\end{equation}
we finally find the gravitoelectric field $\mathrm{\bf{E}}_{g}$ in
the superconductor:
\begin{equation}
\frac{\mathrm{\bf E}_{g}(x,t)}{g^{\ast }}=\left[ 1-\exp (-\left|
\psi _{0}(x)\right| ^{2}t/\eta )-\frac{\partial }{\partial
t}\left( \frac{\exp (-\left| \psi _{0}(x)\right| ^{2}t/\eta
)}{\eta }\int_{0}^{t}dtJ(x,t)\exp (\left| \psi _{0}(x)\right|
^{2}t/\eta )\right) \right]  \label{37}
\end{equation}

From this formula we can see that for maximizing the effect of the
reduction of the gravitational field in a superconductor it is
necessary to reduce $\eta $ and to have large spatial derivatives
of $\psi _{0}(x)$ and $\gamma (x,t)$. The condition for having a
small value of $\eta $ is that the superconductor has a large
normal-state resistivity and a small diffusion coefficient $D\sim
v_{F}l/3$ (where $v_{F}$ is the Fermi velocity, which is small in
HTCS, and $l$ is the mean free path). Therefore, the effect is
enhanced in `bad' samples with impurities, not in single crystals.

From the experimental viewpoint, the greater are the length and
time scales over which there is a variation of $\mathrm{\bf
E}_{g}$, the easier is the observation of this effect. Actually,
we started from non-dimensional equations and therefore the length
and time scales are determined by $\lambda(T)$ and
$\tau=\lambda^{2}(T)/D$, which should therefore be as large as
possible. In this sense, some new materials with very large
$\lambda(T)$ \cite{refe13a} could be interesting for the study of
this effect. Moreover, as clearly seen in eq.~\ref{37}, the
relaxation time is inversely proportional to $|\psi_{0}(x)|^{2}$.
As a result, $\psi_{0}(x)$ must be as small as possible, and this
implies that $\kappa$ is small (see eq.\ref{29}). Then,
$\lambda(T)$ and $\xi(T)$ must be both large.

Up to now we have dealt with the expression of $\beta(x,t)$  as a
function of $\gamma(x,t)$. Actually, to obtain an explicit
expression for {\bf E}$_{g}$ we have to solve the equation for
$\gamma(x,t)$ (eq.~\ref{30}). This is a difficult task which can
be undertaken only in a numerical way. Nevertheless, if one puts
$\psi_{0}(x,t)\simeq 1$ (good approximation in the case of
YBa$_{2}$Cu$_{3}$O$_{7}$ (YBCO), in which $\kappa $=94.4), one can
find the simple approximate solution:
\begin{equation}
\gamma (x,t)=i\gamma _{0}(x)+i\sum_{m=1}^{+\infty }A_{m}\sin (n\pi
x/L)\exp \left[ -(m^{2}\pi ^{2}/L^{2}+2\kappa ^{2})t\right]
\label{38}
\end{equation}
where
\begin{equation}
\gamma _{0}(x)=x/(2\kappa ^{2})\cdot \left\{ 1-\left[ \sinh
(\sqrt{2}\kappa (L-x))+\sinh (\sqrt{2}\kappa x)\right] /\sinh
(\sqrt{2}\kappa L)\right\} \label{39}
\end{equation}

\begin{equation}
A_{m}=\frac{1}{L}\int_{0}^{L}\gamma _{0}(x)\sin (m\pi x/L)
\mathrm{d}x=\frac{L}{2m\pi \kappa ^{2}}\left\{
(-1)^{m}-\frac{\left[ A_{1m}+A_{2m}\right] }{m\pi \left[ \left(
\sqrt{2}\kappa L/(m\pi )\right) ^{2}+1\right] }\right\}
\end{equation}
and
\begin{equation}
A_{1m}=m\pi \left[ 1-(-1)^{m}\cosh (\sqrt{2}\kappa L)\right]
+2\frac{(-1)^{m}\sqrt{2}\kappa L\sinh (\sqrt{2}\kappa L)}{\left[
\left( \sqrt{2}\kappa L/(m\pi )\right) ^{2}+1\right]}
\end{equation}
\begin{equation}
\hspace{-5mm}A_{2m}=\left[ \frac{1-\cosh (\sqrt{2}\kappa L)}{\sinh
(\sqrt{2}\kappa L)}\right] \left\{ \frac{2\sqrt{2}\kappa L\left[
(-1)^{m}\cosh (\sqrt{2}\kappa L)-1\right]}{(m\pi )\left[ \left(
\sqrt{2}\kappa L/(m\pi )\right) ^{2}+1\right] }-m\pi (-1)^{m}\sinh
(\sqrt{2}\kappa L)\right\}\label{A2m}
\end{equation}

By inserting eq.~\ref{39} in eqs.~\ref{34},\ref{35} and taking
into account eq.~\ref{29}, the gravitoelectric field {\bf
E}$_{g}$ becomes:
\begin{equation}
\mathrm{E}_{g}(x,t)=g^{\ast }\left[ 1-\exp (-\left| \psi
_{0}(x)\right| ^{2}t/\eta
)\left( 1-\frac{J_{0}(x)}{\eta }\right) +\sum_{m=1}^{+\infty }\frac{%
A_{m}B_{m}(x)C_{m}(x,t)}{\eta }\right]  \label{41}
\end{equation}
where
\begin{equation}
J_{0}(x)=\frac{1}{2\kappa ^{2}}\left[ \psi _{0}(x)\frac{\partial }{\partial x%
}\gamma _{0}(x)-\gamma _{0}(x)\frac{\partial }{\partial x}\psi
_{0}(x)\right], \label{42}
\end{equation}
\begin{equation}
B_{m}(x)=(m\pi /L)\psi _{0}(x)\cos (m\pi x/L)-\sin (m\pi
x/L)\frac{\partial }{\partial x}\psi _{0}(x),  \label{43}
\end{equation}
\begin{equation}
C_{m}(x,t)=\frac{(m^{2}\pi ^{2}/L^{2}+2\kappa ^{2})\exp \left[
-(m^{2}\pi ^{2}/L^{2}+2\kappa ^{2})t\right] -\left( \left| \psi
_{0}(x)\right| ^{2}/\eta \right) \exp (-\left| \psi _{0}(x)\right|
^{2}t/\eta )}{\left( \left| \psi _{0}(x)\right| ^{2}/\eta \right)
-(m^{2}\pi ^{2}/L^{2}+2\kappa ^{2})}.  \label{44}
\end{equation}

Note that making the very drastic approximation
\begin{equation}
\gamma (x)\simeq ix/(2\kappa ^{2})  \label{45}
\end{equation}
leads to the apparently draft result
\begin{equation}
\mathrm{E}_{g}(x,t)=g^{\ast }\left[ 1-\exp (-\left| \psi
_{0}(x)\right| ^{2}t/\eta )\left( 1-\frac{J_{00}(x)}{\eta }\right)
\right] \label{46}
\end{equation}
where
\begin{equation}
J_{00}(x)=\frac{1}{2\kappa ^{2}}\left[ \psi
_{0}(x)-x\frac{\partial \psi _{0}(x)}{\partial x}\right]
\label{47}
\end{equation}

In spite of its crudeness, this approximate solution
(eq.~\ref{46}) in the case of YBCO gives the same results of the
exact solution (eq.~\ref{41}). Moreover, nothing changes
significantly if one neglects the finite size of the
superconductor and uses $\psi _{0}(x)=\tanh \left( \kappa
x/\sqrt{2}\right)$ instead of eq.~\ref{29}. In the case of YBCO
the variation of the gravitoelectric field E$_{g}$ in time and
space is shown in Figure 1a and 1b. It is easily seen that this
effect is almost independent on the spatial coordinate.

The results in the case of Pb are reported in Figure 2a and 2b,
which clearly show that, due to the very small value of $\kappa$,
the reduction is greater near the surface. In this case, moreover,
some approximations made in the case of YBCO are no longer
allowed. For example, for small values of $L$ the simplified
relation (\ref{45}) is not valid. When $\kappa$ is small, in fact,
the length $L$ plays an important role. In particular, if $L$ is
small the effect is remarkably enhanced, as shown in Figure 3. In
the same condition, a maximum of the effect (and therefore a
minimum of E$_{g}$) can occur at $t\neq 0$, as can be seen in the
same Figure.

In conclusion, for YBCO the shielding effect decays with a
relaxation time $t_{\mathrm{surf}}\simeq $ $0.25$ $\tau =8.5$
$10^{-11}$ s near the surface ($x=0.01 \lambda =3.3\cdot10^{-9}$
m) and $t_{\mathrm{int}}\simeq 0.1\tau =3.4\cdot10^{-11}$s in the
interior of the sample ($x=\lambda=3.3\cdot10^{-7}$m).

In the case of Pb, the same quantities take the values $t_{\mathrm{surf}}%
\simeq 3\cdot10^{7}\tau =1.8\cdot10^{-8}$s ($x=\lambda
=7.8\cdot10^{-8}$m) and $t_{\mathrm{int}}\simeq5\cdot10^{5}\tau
=3.1\cdot10^{-9}$s ($x=100\lambda =7.8\cdot10^{-6}$m) with
$L\gg\lambda$.

Table~1 reports the values of the parameters of YBCO and Pb,
calculated at a temperature $T$ such that $(T-T_{c})/T_{c}$ is the
same in the two materials.
\begin{center}
\begin{tabular}{|c c|c|c c|}
\multicolumn{5}{@{}c @{}}{\bf{Table 1}} \\ \hline & & YBCO & Pb &
\\ \hline \hline \V &$T_{c}$ & 89~K & 7.2~K & \D
\\ \hline \V & $T$ & 77~K & 6.3~K & \D
\\ \hline \V & $\lambda(T)$ & 3.3$\cdot10^{-7}$~m &
7.8$\cdot10^{-8}$~m& \D
\\ \hline \V & $\xi(T)$ & 3.6$\cdot10^{-9}$~m & 1.7$\cdot10^{-7}$~m& \D \\
\hline \V & $\sigma^{-1}$ & 4$\cdot10^{-7}$~$\Omega$m($T=90$K) & 2.5$\cdot%
10^{-9}$~$\Omega$m($T=15$K) & \D \\ \hline \V & $H_{c}(T)$ & 0.2~T
& 0.018~T & \D
\\ \hline \V & $\tau(T)$  & 3.4$\cdot10^{-10}$~s &
6.1$\cdot10^{-15}$~s& \D
\\ \hline \V & $D$ & 3.2$\cdot10^{-4}$~m$^{2}$/s & 1~m$^{2}$/s & \D \\
\hline \V & l & 6$\cdot10^{-9}$~m & 1.7$\cdot10^{-6}$m & \D
\\ \hline \V & $v_{F}$ & 1.6$\cdot10^{5}$~m/s &
1.83$\cdot10^{6}$~m/s& \D
\\ \hline \V & $\kappa$  & 94.4 & 4.8$\cdot10^{-1}$ & \D \\ \hline \V & $\eta$  &
1.27$\cdot10^{-2}$ & 6.6$\cdot10^{3}$& \D \\ \hline
\end{tabular}
\end{center}

Tables~2.a and 2.b summarize the variation of the fundamental
quantities with the temperature~\cite{refe12a}: 

\begin{center}
\begin{tabular}{|c c|c|c|c c|}
\multicolumn{6}{@{}c @{}}{\bf{Table 2.a}} \\ \hline & YBCO &
$\lambda$ & $\tau$ & $g^{\ast }$ & \\ \hline \hline \V &%
$T=0$K & 1.7$\cdot10^{-7}$~m & 9.03$\cdot10^{-11}$s & 2.6%
$\cdot10^{-12}$ & \D \\ \hline \V & $T=70$K & 2.6$\cdot10^{-7}$~m & 2.1$\cdot10^{-10}$s & 9.8%
$\cdot10^{-12}$ & \D \\  \hline \V &$T=77$K & 3.3$\cdot10^{-7}$~m & 3.4$\cdot10^{-10}$s & 2%
$\cdot10^{-11}$ & \D\\ \hline \V & $T=87$K & 8$\cdot10^{-7}$~m & 2$\cdot10^{-9}$s & 2.8%
$\cdot10^{-7}$ & \D\\ \hline
\end{tabular}
\begin{tabular}{|c c|c|c|c c|}
\multicolumn{6}{@{}c @{}}{\bf{Table 2.b}} \\ \hline & Pb &
$\lambda$ & $\tau$ & $g^{\ast
}$ & \\ \hline \hline \V &$T=0$K & 3.90$\cdot10^{-8}$~m & 1.5$\cdot10^{-15}$s & 1%
$\cdot10^{-17}$ \D & \\ \hline \V & $T=4.20$K & 4.3$\cdot10^{-8}$~m & 1.8$\cdot10^{-15}$s & 1.4%
$\cdot10^{-17}$\D & \\ \hline \V & $T=6.26$K & 7.8$\cdot10^{-8}$~m & 6.1$\cdot10^{-15}$s & 8.2%
$\cdot10^{-17}$ \D & \\ \hline \V &$T=7.10$K & 2.3$\cdot10^{-7}$~m & 5.3$\cdot10^{-14}$s & 2.2%
$\cdot10^{-15}$\D & \\ \hline
\end{tabular}
\end{center}

\vspace{3mm} It is clearly seen that $\lambda $ and $\tau $ grow
with the temperature, so that one could think that the effect is
maximum when the temperature is very close to $T_{c}$. However,
this is true only for low-$T_{c}$ superconductors because in
high-$T_{c}$ superconductors (HTSC) fluctuations are of primary
importance for some Kelvin around $T_{c}$. The presence of these
opposite contributions makes it possible that a temperature
$T_{\max }<T_{c}$ exist, at which the effect is maximum.

In all cases, the time constant $T_{int}$ is very small, and this
makes the experimental observation rather difficult. Here I
suggest to use pulsed magnetic fields to destroy and restore the
superconductivity within a time interval of the order of
$T_{int}$.

The main conclusion of this work is that the reduction of the
gravitational field in a superconductor, if it exists, is a
transient phenomenon and depends strongly on the parameters that
characterize the superconductor.

Note that in this paper I have used a very simplified model. For a
more realistic description, one should take into account some
features of real superconductors, for example: \vspace{-2mm}
\begin{enumerate}
\item The symmetry of the order parameter, which in HTCS can be different
from a pure $s$-wave \cite{refe14a};\vspace{-2mm}
\item The fact that the relaxation constant $\eta$ can be complex \cite{refe14aa};\vspace{-2mm}
\item The high anisotropy and layered structure of HTCS
\cite{refe14aaa};\vspace{-2mm}
\item The effect of superconducting fluctuations, which is very large in
HTCS \cite{refe21a}.
\end{enumerate}
Finally, since in the general solution (eq.~\ref{37}) time and
space derivative of the order parameter are present, I suggest
that this effect could be enhanced: \vspace{-2mm}
\begin{enumerate}
\item by the presence of impurities \cite{refe16a};\vspace{-2mm}
\item by using quickly variable (pulsed) magnetic fields \cite
{refe18a};\vspace{-2mm}
\item by making the superconductor quickly rotate \cite{refe20a};\vspace{-2mm}
\item by using constant or time-dependent electric
fields \cite{refe19a};\vspace{-2mm}
\item by manufacturing a superconductor made of layers of
different materials, or different phases (with different $T_{c}$)
of the same material \cite{refe15aa}.
\end{enumerate}
\vspace{10mm}

\textbf{ACKNOWLEDGEMENTS}

I thank R.S. Gonnelli for stimulating discussions, D. Daghero for
the collaboration and C. Pierbattisti for the irreplaceable help.

\newpage
\textbf{ FIGURE CAPTIONS}
\vspace{10mm}

\noindent\textbf{Fig. 1 (a)} The gravitational field
E$_{g}/g^{\ast}$ as a function of the normalized time and space
for YBCO at $T=77$~K; \textbf{(b)} the gravitational field as a
function of the normalized time for increasing values of $x$:
$x=5\cdot10^{-3}\lambda$, $x=10^{-2}\lambda$ and $x=\lambda$.
\vspace{10mm}

\noindent\textbf{Fig. 2 (a)} The gravitational field
E$_{g}/g^{\ast}$ as a function of the normalized time and space
for Pb at $T=6.3$~K; \textbf{(b)} the gravitational field as a
function of the normalized time for increasing values of $x$:
$x=\lambda$, $x=2\lambda$ and $x=10\lambda$. \vspace{10mm}

\noindent \textbf{Fig. 3} The gravitational field E$_{g}/g^{\ast}$
as a function of the normalized time in the case of Pb, with
$L$=6, 8 and 10 and x=4. The maximum of the effect is evident.


\begin{figure}
\includegraphics[keepaspectratio,height=\textheight]{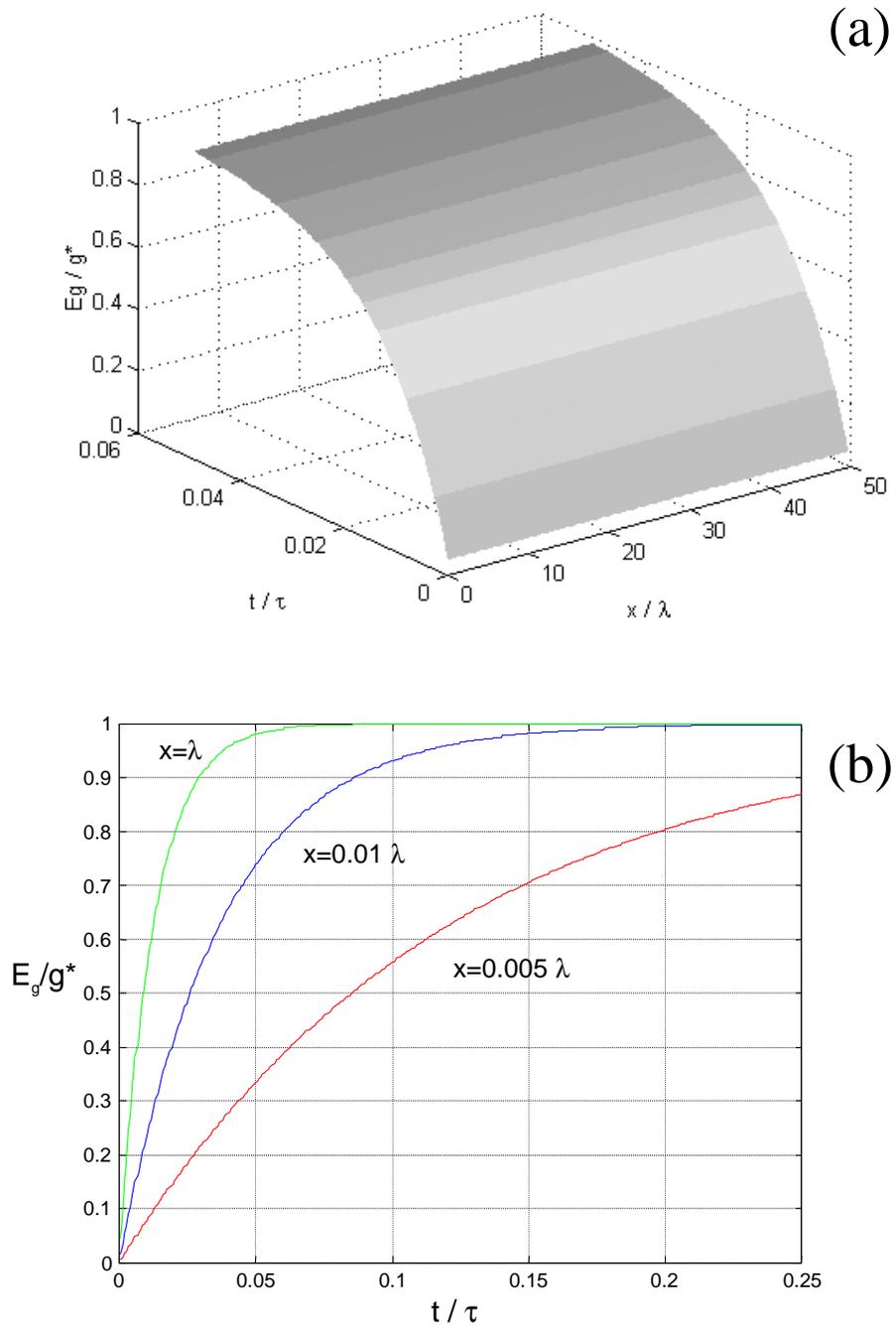}
\caption{G. A. Ummarino, \emph{Possible alterations of the
gravitational field in a superconductor}}
\end{figure}

\begin{figure}
\includegraphics[keepaspectratio,height=\textheight]{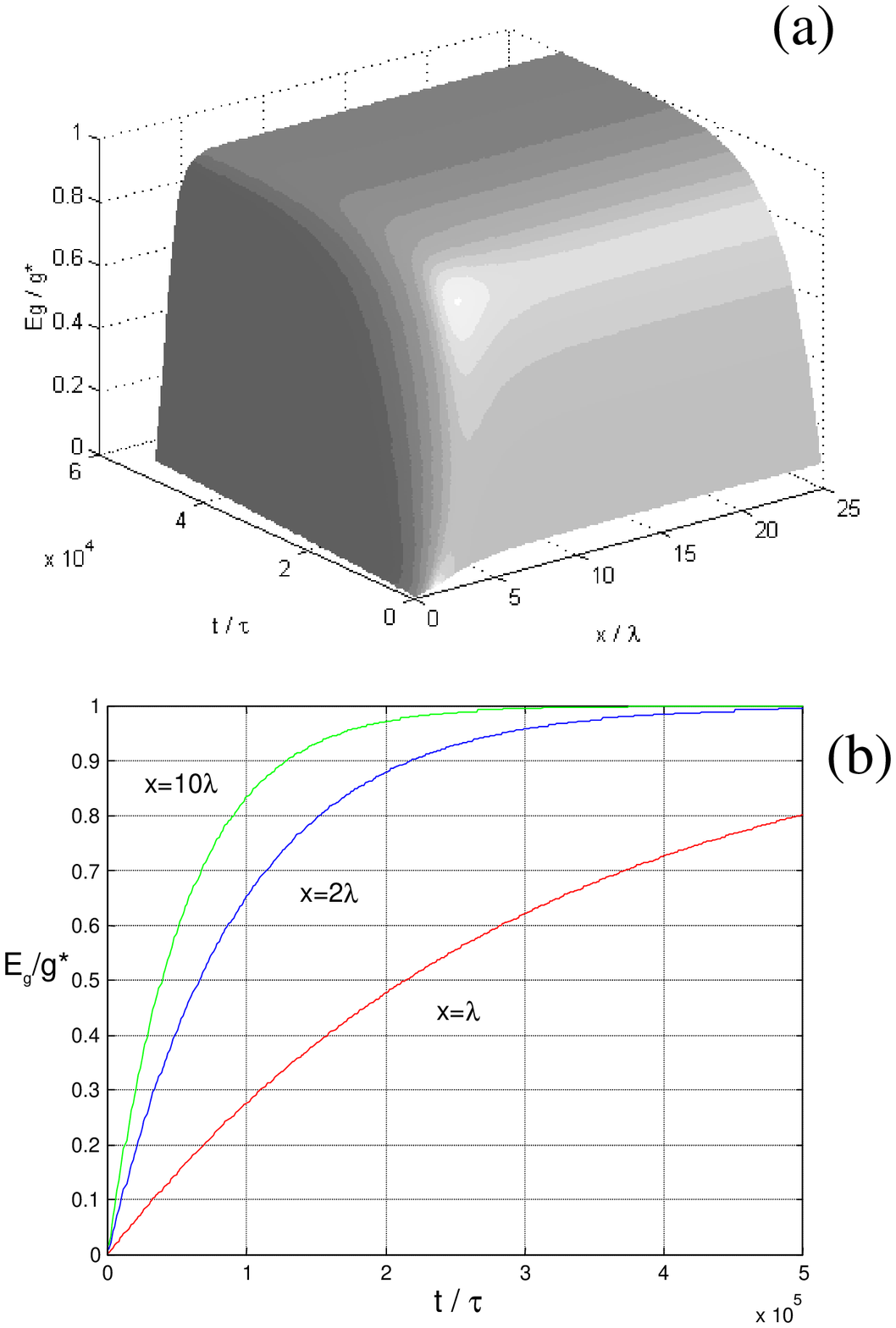}
\caption{G. A. Ummarino, \emph{Possible alterations of the
gravitational field in a superconductor}}
\end{figure}

\begin{figure}
\vspace{30mm}\includegraphics[keepaspectratio,width=\textwidth]{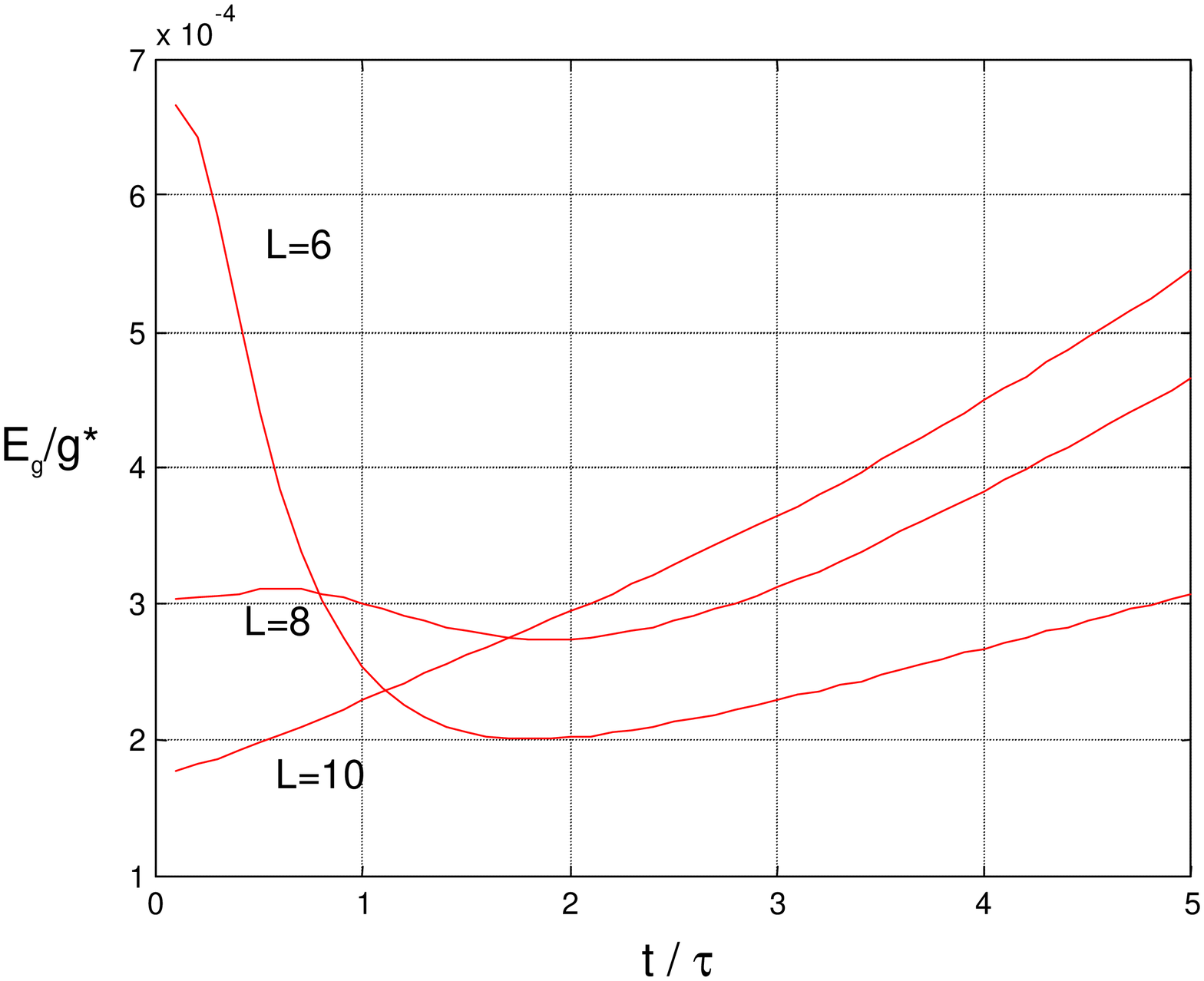}
\vspace{30mm} \caption{G. A. Ummarino, \emph{Possible alterations
of the gravitational field in a superconductor}}
\end{figure}


\begin{thebibliography}{1}

\bibitem{refe0a} B.S. De Witt, Phys. Rev. Lett., {\bf 16} (1966) 1092; G.
Papini, Phys. Lett. {\bf 54A }(1967) 32; S.B. Felch, J. Tate, B.
Cabrera and J.T. Anderson, Phys. Rev. B {\bf 31} (1985) 7006.

\bibitem{refe0b} J. Anadan, Class. Quantum Grav. {\bf 11} (1994) A23; J. Anadan, Il
Nuovo Cimento {\bf 53A} (1979) 21; J. Anadan, Phys. Lett. {\bf
105A} (1984) 280; D.K. Ross, J. Phys. A: Math. Gen. {\bf 16}
(1983) 1331; H. Hirakawa, Phys. Lett. {\bf 53A} (1975) 395.

\bibitem{refe0c} R.G. Rystephanick, Can. Journ. of Phys., {\bf 51} (1973) 789;
Huei Peng, Gordon Lind and Y. S. Chin, General Relativity and
Gravitation {\bf 23} (1991) 1231; C. Ciubotariu and M. Agop, Gen.
Rel. and Grav. {\bf 28} (1996) 405.

\bibitem{refe0d} M. Agop, C.G. Buzea, V. Griga, C. Ciubotariu, C. Buzea, C. Sta, D.
Jatomir, Australian Journ. of Phys. {\bf 49} (1996) 1063; R.P.
Lano, hep-th/9603077; R.P. Lano, gr-qc/961123; O. Yu. Dinariev,
A.G. Mosolov, Sov. Phys. Journ. {\bf 32} (1989) 315; O. Yu.
Dinariev, A.G. Mosolov, Sov. Phys. {\bf 32} (1987) 576; O. Yu.
Dinariev, A.G. Mosolov, Sov. Phys. {\bf 32} (1987) 843; N Sal'e,
R. Tepfer, Sov. Phys. Journ. {\bf 25} (1982) 152; N Sal'e, R.
Tepfer, Exp. Tech. der Phys. {\bf 28} (1980) 13; I.S. Minasyan,
Dokl. Akad. Nauk SSSR, {\bf 228} (1976) 576; F. Rothem, Helv.
Phys. Acta, {\bf 41} (1968) 591.

\bibitem{refe1a}  N. Li, D.G. Torr, Phys. Rev. D {\bf 43} (1991) 457; H.
Peng, D.G. Torr, E.K. Hu, B. Peng, Phys. Rev. B {\bf 43} (1991) 2700; N. Li,
D.G. Torr, Phys. Rev. B {\bf 46} (1992) 5489; D.G. Torr, N. Li, Found. Phys.
Lett., {\bf 6} (1993) 371.

\bibitem{refe2a}  E. Podkletnov and R. Nieminem, Physica C {\bf 203} (1992)
441; E. E. Podkletnov cond-mat/9701074.

\bibitem{refe3a}  Ning Li, David Noever, Tony Robertson, Ron Koczor, Whitt
Brantley, Physica C {\bf 281} (1997) 260.

\bibitem{refe4a}  M. Bull, M. De Podesta, Physica C {\bf 253} (1995) 199;
C.S. Unnikrishnan, Physica C {\bf 266} (1996) 133.

\bibitem{refe5a}  G. Modanese, Europhys. Lett., {\bf 35, (}1996{\bf ) }413%
{\bf ; }G. Modanese, Phys. Rev. D {\bf 54} (1996) 5002.

\bibitem{refe6a}  M. Agop, C. Gh. Buzea, P. Nica, Physica C {\bf 15} (2000)
120; M. Agop, P. D. Ioannou, F. Diaconu, Progress of Theoretical
Physics {\bf 104} (2000) 733.

\bibitem{refe7aa}  C.W. Misner, K.S. Thorne and J.A. Wheeler, (1973) {\it %
Gravitation} (W.H. Freeman and Company).
\bibitem{refe7a} V.B. Braginsky, C.M. Caves, K.S. Thorne, Phys. Rev. D {\bf 15}, (1977) 2047 ;
 Huei Peng, General Relativity and Gravitation {\bf 15} (1983) 725; {\bf 22} (1990) 609; C.J. de Matos, R.E. Becker, gr-qc/9908001.


\bibitem{refe8a}  Qi Tang, S. Wang, Physica D {\bf 88} (1995) 139; Fang-Hua
Lin and Qiang Du, Siam J. Math. Anal., 28 (1997) 1265.

\bibitem{refe9a}  Salman Ullah and Alan T. Dorsey, Phys. Rev. B {\bf 44}
(1991) 262; M. Ghinovker, I. Shapiro, B. Ya Shapiro, Phys. Rev. B
{\bf 59} (1999) 9514; N.B. Kopnin and E.V. Thuneberg, Phys. Rev.
Lett. {\bf 83} (1999) 119; J. Fleckinger-Pell\`{e}, H.G. Kaper and
P. Takac, Nonlin. Anal. Meth. and Appl., {\bf 32} (1998) 647.

\bibitem{refe10a}  Quiang Du and Paul Gray, Siam J. Appl. Math., {\bf 56}
(1996) 1060; Andrew J. Dolgert, cond-mat/9907213.


\bibitem{refe11a}  Bahram Mashhoon, Ho Jung Paik, Clifford M. Will, Phys.
Rev. D {\bf 39} (1989) 2825; A. Ljubicic and B.A. Logan, Phys.
Lett. {\bf 172A} (1992) 3.

\bibitem{refe12a}  D.R. Harshman and A.P. Mills, Jr., Phys. Rev. B {\bf 45}
10684 (1992); C.P. Poole Jr., H.A. Farach, R.J. Creswick, (1995) {\it %
Superconductivity} (Academic Press, Inc.).

\bibitem{refe13a}  J. Dow, H.A. Blackstead and D.R. Harshman, {\it %
Proceedings of International Conference Stripes 2000}, (Rome
2000).

\bibitem{refe14a}  Q.-H. Wang and Z.D. Wang, Physica C {\bf 282-287} (1997)
1967; Jian-xin Zhu, Wonkee Kim, C.S. Ting, Chia-Ren Hu, Phys. Rev. B {\bf 58}
(1992) 15020.

\bibitem{refe14aa}  A.T. Dorsey, Phys. Rev. B {\bf 46} (1993) 8376.

\bibitem{refe14aaa}  Yu. M. Ivanchenko, Phys. Rev. B {\bf 48} (1993) 15966;
D.E. Prober, M.R. Beasley and R.E. Schwall, Phys. Rev. B {\bf 15} (1977)
5245; M. Tinkham, Physica C {\bf 235-240} (1994) 3; D. Feinberg, S.
Theodorakis, A.M. Ettouhami, Phys. Rev. B {\bf 49} (1994) 6285.

\bibitem{refe21a}  A. Varlamov, G. Balestrino, E. Milani, D. Livanov, Adv.
in Phys. {\bf 48 }(1998) 655.

\bibitem{refe16a}  Jerome J. Krempasky and Richard S. Thomson, Phys. Rev. B
{\bf 32} (1985) 2965.

\bibitem{refe18a}  H.G. Kaper, P. Takac, Nonlinearity
{\bf 11} (1998) 291; A.T. Dorsey, Phys. Rev. B {\bf 51} (1995)
15329;A. Gurevich, E.H. Brandt, Phys. Rev. B {\bf 55} (1997)
12706; A.A. Kozhevnikov, Phys. Lett. A {\bf 202} (1995) 343; V.K.
Vlasko-Vlasov, U. Welp, GW. Crabtree, D. Gunter, V.V. Kabanov,
V.I. Nikitenko, L.M. Paulius, Phys. Rev. B {\bf 58} (1998) 3446.

\bibitem{refe20a}  Mario Liu, Phys. Rev. Lett. {\bf 81} (1998) 3223; Brandon
Carter, David Langlois, Reinhard Prix, Phys. Rev. B {\bf 62}
(2000) 9740; Brandon Carter, David Langlois, Reinhard Prix, Phys.
Rev. B {\bf 62} (2000) 9748; J. A. Geurst, Physica B {\bf 101}
(1980) 82; J. A. Geurst, H. van Beelen, Physica C {\bf 208} (1993)
43, Yimin Jiang, Mario Liu, cond-mat/0010260.

\bibitem{refe19a} P. Konsin and B. Sorkin, Phys. Rev. B {\bf 58} (1998)
5795 and references therein.

\bibitem{refe15aa}  Alan T. Dorsey, Annal of Physics, {\bf 233} (1994) 248;
R. Lipowsky and W. Speth, Phys. Rev. B {\bf 28} (1983) 3983.

\end{thebibliography}
\end{document}